\documentclass[a4paper,11pt]{article}

\usepackage{jheppub} 

\usepackage[T1]{fontenc} 

\usepackage{amsmath,amsfonts,amssymb,amsthm}
\usepackage{hyperref}
\hypersetup{
    pdftex,
    pdfstartview=FitH,
    bookmarksnumbered=true,
    pagebackref,
    colorlinks,
    breaklinks,
    urlcolor=RoyalBlue,
    linkcolor=RoyalBlue,
    citecolor=NavyBlue
}
\usepackage{graphicx}
\DeclareGraphicsExtensions{.pdf,.png,.jpg,.mps}
\usepackage{booktabs}
\usepackage[dvipsnames]{xcolor}
\usepackage{tikz}
\usepackage{shuffle}
\usepackage{cleveref}
\usetikzlibrary{calc,matrix,decorations.pathmorphing,decorations.markings,arrows,positioning,intersections,mindmap,backgrounds}

\usepackage{amsmath,amsthm,amsfonts,amssymb,amscd,mathtools
}
\usepackage{pifont}

\newcommand{\zb}{\bar{z}}

\newcommand{\bs}{\mathbf{s}}
\newcommand{\bt}{\mathbf{t}}
\newcommand{\bu}{\mathbf{u}}

\newcommand{\tils}{\tilde{s}}
\newcommand{\tilt}{\tilde{t}}

\title{\LARGE{Bootstrap AdS Veneziano Amplitude with Arbitrary Kaluza-Klein Modes}}

\author[]{Bo Wang$^{a,b}$,}

\affiliation[]{$^{a}$Zhejiang Institute of Modern Physics, School of Physics, Zhejiang University, \\Hangzhou, Zhejiang 310058, China }
\affiliation[]{$^{b}$Joint Center for Quanta-to-Cosmos Physics, Zhejiang University,
\\Hangzhou, Zhejiang 310058, China}

\abstract{We present a derivation of the first curvature correction to the AdS Veneziano amplitude for arbitrary Kaluza-Klein (KK) modes, using a bootstrap approach based on the world-sheet representation and AdS$\times$S formalism. Our results establish a universal formula for the first order curvature correction without considering any low-lying KK configurations. We give new predictions for Wilson coefficients in the low-energy expansion. In the high-energy regime, the amplitude exhibits a universal exponent independent of the external KK charges, providing a coherent picture of AdS stringy amplitudes in different backgrounds.
}

\emailAdd{b\_w@zju.edu.cn}

\date{\today}

\begin{document} \maketitle

\newpage

\section{Introduction}

The AdS Veneziano amplitude describes open-string scattering in AdS backgrounds and arises naturally in holographic models on an AdS$_5\times$S$^3$ background\footnote{One can construct this theory by inserting $N$ probe D3-branes near a F-theory singularity \cite{Sen:1996vd}, see also \cite{Fayyazuddin:1998fb,Aharony:1998xz} for other configurations. The dual 4d $\mathcal{N}=2$ SCFT is realized in a $USp(2N)$ gauge theory with flavor $SO(8)$ group.}. In the AdS/CFT correspondence, such amplitudes are dual to four-point correlators of half-BPS operators in a 4d $\mathcal{N}=2$ SCFT, including external states with nontrivial Kaluza–Klein (KK) charges. An important motivation for studying \textit{string} amplitudes directly in AdS is the expectation that any holographic CFT with a curved string dual should possess a world-sheet description. The difficulty is the presence of Ramond-Ramond (RR) flux in a curved background, which is extremely hard to handle by the traditional string theory approach. 

Recent progress on the AdS Virasoro–Shapiro amplitude \cite{Alday:2023jdk,Alday:2023mvu,Fardelli:2023fyq,Wang:2025pjo} obtained from a bootstrap perspective has uncovered a concrete AdS world-sheet construction. Its predictions for anomalous dimensions and OPE coefficients agree strikingly with integrability and localization results \cite{Pestun:2016zxk,Beisert:2010jr,Gromov:2013pga}. On the open-string side, an AdS analogue of the Veneziano amplitude has been identified, but existing results mostly concern the lowest KK sector \cite{Alday:2024yax,Alday:2024ksp}. These developments suggest that world-sheet methods in AdS not only reproduce the super-gluon amplitude; but also organize more stringy effects in a constraining way.

The main idea of this work is to extend this program from the lowest external states to \textit{arbitrary} external KK modes for the open-string amplitude in an AdS background. There are both conceptual and practical reasons to do so. Conceptually, turning on KK charges excites a richer tower of string states in the intermediate channel and probes additional OPE data; on the low-energy side, it provides the information about the higher-derivative couplings in the AdS effective action \cite{Glew:2023wik}. Practically, however, superconformal block expansions become increasingly cumbersome once KK structure is included due to the mixing among nearly degenerate operators and the complexity of R-symmetry structures. 

Prior approaches effectively navigated this by fixing a handful of low-lying KK configurations and then uplifting the result to general charges \cite{Wang:2025pjo}. Our aim here is to design a bootstrap procedure that \textit{directly} targets arbitrary KK modes and closes without requiring any seed data from special low-lying cases. By doing so, we find the explicit structure of exchanged averaged twists and extract a broad set of usable averaged OPE data at each string level, valid for arbitrary KK modes.

Physically, we work in the large $N$ limit and the AdS radius is parametrically large compared to the string length. The single expansion parameter is the AdS curvature,
\begin{equation}
\lambda^{-1/2}=\frac{\alpha'}{R^2} \; \ll \; 1\,,
\end{equation}
so the discussion applies when one thinks of taking $R$ large at fixed $\alpha'$. In this limit, the first curvature correction $\mathcal{A}^{(1)}\sim \lambda^{-\frac{1}{2}}$ captures the leading effect from AdS bulk. On the CFT side, the exchanged intermediate states in the $s$-channel are \textit{massive string} excitations with twist scaling as
\begin{equation}
\tau \;=\; \sqrt{\delta}\,\lambda^{1/4} \;+\; \tau_1 \;+\; \tau_2\,\lambda^{-1/4} \;+\cdots,
\end{equation}
where $\delta=1,2,\dots$ labels the string level and $\ell=0,1,\dots,\delta-1$ the spin. These towers generate physical poles at $S=\delta$ with finite spin support. After an appropriate Borel transform, the $S$ poles of the AdS amplitude are in one-to-one correspondence with logarithmic singularities of the world-sheet integrand near s-channel OPE $z\to 0$, so the AdS curvature expansion can be organized directly in terms of the world-sheet function.

We develop a systematic bootstrap framework based on a world-sheet ansatz in AdS$\times$S space. The heart of this paper is the following methodological pipeline, which we find to be \textit{overconstrained} at the first curvature correction,
\begin{equation}
\boxed{\scriptsize
\text{Super-block decomposition}
\;\xRightarrow[\text{Large twist}]{\text{Borel transform}}\;
\text{AdS}\times\text{S on world-sheet}
\;\Rightarrow\;
\text{unique } \mathcal{A}^{(1)} \text{ for any KK.}}
\end{equation}
Concretely, we start from the superblock decomposition, perform a large-twist expansion and a Borel transform \cite{Alday:2022xwz} that isolates the flat-space pole structure, and then construct a world-sheet integrand on the disk. We organize the integrand in terms of a finite basis of single-valued multiple polylogarithms (SVMPLs) on the real line \cite{Alday:2024ksp}, multiplied by suitable rational prefactors in $(S,T,U)$ and in the spherical Mellin variables $(n_s,n_t,n_u)$ that encode the internal data \cite{Aprile:2020luw}. Matching the residues at the physical poles $S=\delta$ fixes all unknown coefficients, yielding a \textit{unique} consistent answer for $\mathcal{A}^{(1)}(S,T)$. We expect this pipeline and the resulting uniqueness at first curvature order can be extended to a broad class of supersymmetric holographic backgrounds \cite{Chester:2024wnb,Chester:2024esn,Jiang:2025oar}.

This work stands at the intersection of several recent developments. On the one hand, it complements the AdS stringy bootstrap studies by focusing on the open-string sector and demonstrating that similar bootstrap principles apply with arbitrary KK modes. On the other hand, our results provide a new derivation of the amplitude that was previously only indirectly constrained. Unlike effective-action approaches \cite{Glew:2023wik} that rely on an assumed 8d action, our method is based on the AdS/CFT correspondence. We thereby extend previous findings by removing the need for external configurations and by incorporating all external KK modes in a unified framework. After identifying the superblock expansion and the world-sheet ansatz in AdS$\times$S space, all of the undetermined coefficients are fixed, yielding a complete solution. Our final world-sheet formula further provides the infinite sub-leading terms in the low-energy expansion. Finally, we also find intriguing evidence that the high-energy limit of the amplitude is independent of the external KK charges, hinting at a deeper organizing principle in AdS string dynamics.

In the following sections, we set up the AdS$_5\times$ S$^3$ correlator kinematics and Mellin representation using AdS$\times$S formalism. In \cref{sec:worldsheet}, we will construct the world-sheet ansatz for the Veneziano amplitude. In \cref{sec:exactres}, we perform the bootstrap matching to obtain the exact amplitude at the first curvature correction. Next, in \cref{sec:limit} we analyze the low-energy expansion to extract Wilson coefficients and the high-energy limit, whose behavior indicates the universality of AdS string amplitude. Finally, we discuss the implications of our findings and future directions.

\section{Setup and Strategy}\label{sec:setup}

\subsection{Correlator}

We consider type IIB string theory in an $\mathrm{AdS}_5 \times$S$^3$ background, which arises as the near-horizon geometry of D3-branes probing F-theory singularities. The dual field theory is a four-dimensional $\mathcal{N}=2$ superconformal field theory (SCFT), featuring gauge group ${USp}(2N)$ and flavor symmetry group $G_F = {SO}(8)$. The internal space $S^3$ breaks the original $SO(6)$ isometry of S$^5$ to $SU(2)_R \times SU(2)_L \times U(1)_R$, where $SU(2)_R \times U(1)_R$ forms the R-symmetry and $SU(2)_L$ acts as a global symmetry.

The external states are super primaries of half-BPS operators in a $\mathcal{N}=2$ SCFT dual to KK modes with angular momentum $k$ on the S$^3$. They can be interpreted as super-gluons in the AdS bulk\footnote{More precisely, the super-gluon states correspond to single-particle excitations in AdS, which are not dual to individual single-trace operators, but to specific linear combinations thereof. These combinations are determined by the mixing structure of the SCFT, see the detailed discussion in appendix C of \cite{Huang:2023ppy}.}. These operators transform under non-trivial representations of both $SU(2)_R$ and $SU(2)_L$, with their polarization information encoded in the $SU(2)$ spinors $v^a$ and $\bar{v}^{\dot{a}}$ ($a, \dot{a} = 1,2$). The general structure of such operators is
\begin{equation}\label{eq:vvb}
\mathcal{O}^I_k(x; v, \bar{v}) = \mathcal{O}^{I;a_1 \cdots a_k;\dot{a}_1 \cdots \dot{a}_{k-2}}(x)\, v_{a_1} \cdots v_{a_k} \bar{v}_{\dot{a}_1} \cdots \bar{v}_{\dot{a}_{k-2}},
\end{equation}
where $I$ labels the flavor index. The label $k$ determines the KK level and conformal dimension $\Delta = k$ of the corresponding operator.

We will focus on the four-point function of such half-BPS operators,
\begin{equation}
    G^{I_1 I_2 I_3 I_4}_{\{p_i\}}\equiv \langle \mathcal{O}^{I_1}_{p_1}(x_1;v_1,\bar{v}_1) 
\mathcal{O}^{I_2}_{p_2}(x_2;v_2,\bar{v}_2) 
\mathcal{O}^{I_3}_{p_3}(x_3;v_3,\bar{v}_3) 
\mathcal{O}^{I_4}_{p_4}(x_4;v_4,\bar{v}_4) \rangle.
\end{equation}
This function depends on the spacetime positions $x_i$, the R-symmetry spinors $v_i$, $\bar{v}_i$, and the external dimensions $p_i$. Due to supersymmetry, this correlator satisfies superconformal Ward identities \cite{Nirschl:2004pa} and admits a decomposition into a free part $G^{I_1 I_2 I_3 I_4}_{\text{free}}$ and an interacting part,
\begin{equation}
    G^{I_1 I_2 I_3 I_4}_{\{p_i\}} = G^{I_1 I_2 I_3 I_4}_{\text{free}} + R_{1234} H^{I_1 I_2 I_3 I_4}_{\{p_i\}}
\end{equation}
where $R_{1234}$ is a universal prefactor fixed by supersymmetry. This expression enjoys a symmetric form \cite{Huang:2024dxr}
\begin{equation}
     R_{1234}=V_{1234} x^2_{13}x^2_{24} + V_{1342} x^2_{14}x^2_{23}+ V_{1423} x^2_{12}x^2_{34} \, , \quad V_{ijkl} = v_{ij} v_{jk} v_{jl} v_{li} \, ,
\end{equation}
where $x_{ij}=x_i-x_j$, $v_{ij} = \epsilon^{a b}v_a v_b$ (similarly for $\bar v$). The reduced correlator ${H}^{I_1 I_2 I_3 I_4}_{\{p_i\}}$ captures all of the dynamical information.

We only consider the genus-0 contribution from the string theory side, so we will limit our discussion to large $N$. In this planar limit, only a particular single-trace structure contributes to the four-point function. Specifically, we focus on the colour-ordered correlator $H_{\{p_i\}}$ in the full correlator,
\begin{equation}\label{eq:color-structure}
H^{I_1 I_2 I_3 I_4}_{\{p_i\}}=\mathrm{Tr}[T^{I_1} T^{I_2} T^{I_3} T^{I_4}] H_{\{p_i\}}+\text{crossing} \, ,
\end{equation}
where $T^I$ are the generators of the flavor group $G_F = {SO}(8)$. This trace ordering corresponds to the color-ordered partial amplitude in string theory, dual to the planar disk diagram on the world-sheet. This choice isolates the single-trace contribution to the correlator and matches the structure of the open string amplitude. Other color structures (such as double-trace terms) are suppressed at large $N$.

At strong 't Hooft coupling $\lambda$, the dynamical part ${H}$ admits an expansion in $\lambda^{-\frac{1}{2}}$
\begin{equation}
    H_{\text{genus-0}} = H_{\text{super-gluon}} + \lambda^{-1} H_2 + \lambda^{-3/2} H_3 + \cdots \, ,
\end{equation}
where $H_{\text{super-gluon}}$ corresponds to the field theory limit contribution calculated in \cite{Alday:2021odx}, and $H_i$ arises from stringy corrections.

\subsection{Generalized Mellin amplitude}

We now turn to the Mellin representation, which provides a powerful framework for analyzing four-point correlation functions. In Mellin space, the analytic structure of the correlator becomes more transparent, closely mirroring the properties of flat-space scattering amplitudes—such as meromorphic dependence on kinematic variables. We further consider the AdS$\times$S Mellin formalism to capture the internal symmetry dependence on S$^3$. 

The AdS$\times$S formalism was originally developed to efficiently organize the internal symmetry structure of holographic correlators involving Kaluza-Klein (KK) modes in theories such as type IIB string theory on $\mathrm{AdS}_5 \times$S$^5$ \cite{Aprile:2020luw} and similar backgrounds\footnote{See also recent work on higher point amplitude bootstrap in AdS$\times$S space \cite{Huang:2024dxr,Fernandes:2025eqe}.}. In such theories, the four-point amplitude of half-BPS operators in the boundary CFT (e.g., $\mathcal{N}=4$ SYM) admits a remarkably compact expression.

Let us construct the AdS Veneziano amplitude as suggested in \cite{Wang:2025pjo}. To handle the internal data, the AdS$\times$S formalism introduces Mellin-like variables $n_{ij}$ defined via a spherical Mellin transform, leading to the so-called \textit{AdS$\times$S transform},
\begin{equation}
H(x^2_{ij},y_{ij}) = \int \mathllap \sum [\text{d} \gamma ] \mathcal{M}(\gamma_{ij} ; n_{ij}) \prod_{i<j} \frac{ \Gamma(\gamma_{ij})}{\Gamma(n_{ij}+1)} \frac{y_{ij}^{n_{ij}}}{(x_{ij}^2)^{\gamma_{ij}}} \, ,
\end{equation}
where $y_{ij}={v}_{ij}\bar{v}_{ij}$ and the variables $\gamma_{ij}$ and $n_{ij}$ are constrained by
\begin{align}
\sum_{i} \gamma_{ij}=0& \, , \qquad  \gamma_{ij} = \gamma_{ji} \, , \qquad \gamma_{ii} = -(p_i + 1) \, , \\
\sum_i n_{ij} = 0& \, , \qquad n_{ij} = n_{ji} \, , \qquad n_{ii} = -(p_i - 2) \, . \label{eq:sumn}
\end{align}
One needs to sum over all integers $n_{ij}$ satisfying the \cref{eq:sumn}. Notice the $\mathcal{N}=2$ supersymmetry shifts $p_i$ to $p_i+1$ for $\gamma_{ij}$ and $p_i$ to $p_i-2$ for $n_{ij}$. The integral measure reads
\begin{equation}
    [\text{d}  \gamma ] = \prod {\frac{\text{d} \gamma_{ij}}{2 \pi i}} .
\end{equation}
An important observation is that certain crossing-symmetric combinations of $\gamma_{ij}$ and $n_{ij}$ appear repeatedly in both super-gluon amplitude and string corrections. We can introduce the following Mellin variables
\begin{align}
& s = \Sigma - \gamma_{12} -\gamma_{34}, &  & t = \Sigma - \gamma_{14} -\gamma_{23} , & & u = \Sigma - \gamma_{13} -\gamma_{24}, &\\
& n_s = n_{12} + n_{34}, &  & n_t = n_{14} + n_{23}, & & n_u = n_{13} + n_{24}, &
\end{align}
subject to the constraint
\begin{equation}
s+t+u= 2\Sigma -2 \, , \qquad n_s + n_t + n_u = \Sigma - 4 \, ,
\end{equation}
where $\Sigma= (p_1+p_2+p_3+p_4)/2$. Then the Mellin amplitude $\mathcal{M}(\gamma_{ij},n_{ij})$ encodes both dynamical information and internal symmetry dependence. These variables $n_s$, $n_t$, and $n_u$ are natural analogs of the Mellin variables $s$, $t$, and $u$ in CFT, and appear in the rational prefactors of the world-sheet integrand. Their crossing properties mirror those of the kinematic variables, making them ideal for constructing a crossing symmetric ansatz.

The use of the AdS$\times$S transform and $n_{ij}$-variables also streamlines the matching to known results for gluon amplitudes in AdS, particularly in the super-gluon limit. For example, the tree-level contribution to the four-point function of super-gluons derived in \cite{Alday:2021odx} has a remarkably simple expression in the AdS$\times$S representation \cite{Drummond:2022dxd},
\begin{equation}\label{eq:supergluon}
\mathcal{M}_{\text{super-gluon}}(s, t; n_{ij}) = -\frac{1}{2}\frac{1}{({\bf{s}} + 1)( {\bf{t}} + 1)},
\end{equation}
with $\bs = (s+n_s-\Sigma)/2 $, and similarly for $\bt$ and $\bu$, showing a clear dependence on $n_{ij}$ in parallel with spacetime variables. This structural simplicity extends to stringy corrections. The low-energy expansion of the reduced amplitude $\mathcal{M}(s,t;n_{ij})$ reads
\begin{equation}
    \mathcal{M}(s,t;n_{ij})_{\text{genus-0}} = \mathcal{M}_{\text{super-gluon}} + \lambda^{-1} \mathcal{M}_2 + \lambda^{-3/2} \mathcal{M}_3 + \cdots.
\end{equation}
The first derivative term (of order $\lambda^{-1}$) contributes a constant term $2^4\,(\Sigma-2)2\zeta(2)$ to $\mathcal{M}(s,t;n_{ij})$. However, higher-string corrections generally depend more complicatedly on the external dimensions $p_i$. In our case, this dependence can be captured entirely by the combinations $n_s, n_t, n_u$.

\subsection{Superconformal block decomposition}

The Mellin amplitude can be decomposed into superconformal blocks, labeled by spin $\ell$ and twist $\tau = \Delta - \ell$, written as a sum over Mack polynomials,
\begin{equation} \label{eq:decomMack}
     \mathcal{M}(s,t) \sim \sum_{\mathcal{O}} \sum_{m=0}^{\infty}   \frac{ C_{\mathcal{O}} }{s - \tau - 2m } \, \mathcal{Q}^{\tau+2}_{\ell,m}(u-p_1-p_4),
\end{equation}
where $\mathcal{Q}$ are Mack polynomials associated to spin-$\ell$ exchanges with twist $\tau$ for $m\in \mathbb{N}_0$. This decomposition is analogous to a partial wave expansion in flat space. 

A given exchanged operator with twist $\tau$ and spin $\ell$ receives contributions from several $R$-symmetry representations, and the superconformal block decomposition sums over all of them. In the large–twist regime, however, the splittings among these representations are parametrically suppressed, so their twists become degenerate. We may therefore trade the representation-by-representation sum for averaged data and an \emph{effective} single–twist sector at each string level $\delta$. Equivalently, the OPE coefficients implicitly carry the S$^3$ information, and after summing over the spherical labels one obtains averaged combinations that enter the residues. In this sense, the contributions to the AdS Veneziano amplitude are effectively controlled by a single \emph{averaged} twist per string level, reflecting the large anomalous dimensions of stringy intermediate states at strong coupling. The massive stringy state exchanged in the middle reads
\begin{equation}
    \Delta= m R \left(1 + \mathcal{O}(\lambda^{-\frac{1}{2}}) \right) = \sqrt{\delta}\lambda^{\frac{1}{4}} +  \mathcal{O}(\lambda^{-\frac{1}{4}}) \, ,
\end{equation}
where $m^2 \alpha' =\delta =1,2,\cdots$ label the string mass level. We expect that the OPE data and anomalous dimensions in large $\lambda$ can be written as\footnote{In AdS$\times$S space, the OPE data formally carry the dependence on the compact sphere through the spherical Mellin variables. For a specific KK configuration, one first performs the spherical Mellin transform and then decomposes the result into normal spherical harmonics, which projects onto R-symmetry representations and yields standard OPE coefficients.}
\begin{align} 
    \tau(r;\lambda)&=\sqrt{\delta} \lambda^{\frac{1}{4}}+\tau_1(r)+\tau_2(r)\lambda^{-\frac{1}{4}}+\cdots \label{eq:largetau1}\\
     f(r;\lambda)&=f_0(r)+f_1(r)\lambda^{-\frac{1}{4}}+f_2(r)\lambda^{-\frac{1}{2}}+\cdots \label{eq:largetau2}\\
    C_{\mathcal{O}}(r;\lambda)&=\frac{\pi ^3 \,  \tau ^{2\Sigma-4} 4^{-\Sigma- \ell- \tau+1}} { (\ell+1) \sin \left(\frac{1}{2} \pi  (\tau- p_1-p_2 )\right) \sin \left(\frac{1}{2} \pi  (\tau -p_3 - p_4)\right)} f(r;\lambda) \, . \label{eq:largetau3}
\end{align}

In order to obtain a well-defined amplitude without divergences, we take the flat-space limit via the Borel transform \cite{Penedones:2010ue,Alday:2022xwz},
\begin{equation}\label{eq:Borel}
    \mathcal{A}(S,T)=\frac{\lambda}{8}  \, \Gamma(\Sigma-2) \, \lim_{\lambda\to \infty}\int d\alpha \,  \frac{e^{\alpha} }{\alpha^{\Sigma}} \,  \mathcal{M}\left(\frac{\sqrt{\lambda} S}{2 \alpha} + \frac{2\Sigma-2}{3}, \frac{\sqrt{\lambda} T}{2 \alpha} +  \frac{2\Sigma-2}{3}\right).
\end{equation}
This transformation provides a smooth interpolation from the flat-space string limit to the curved AdS regime, allowing us to bootstrap amplitudes with KK modes.

To proceed from the Mellin decomposition of the correlator to a physical flat-space scattering amplitude, we must perform two key operations following previous discussion, a large twist expansion and a Borel resummation.

At large $\lambda$, the twist $\tau$ of the exchanged stringy states becomes large. We parametrize this large twist limit by introducing a continuous variable $x$, such that the quantum number $m$ scales as $m = x\, \tau^2$. Then the sum over $m$ in the Mellin decomposition becomes an integral over $x$ 
\begin{equation}
    \sum_{m=0}^{\infty} \to \int_{0}^{\infty} dx \, \tau^2 \, .
\end{equation}
This continuous approximation allows us to combine the large $\tau$ behavior of the Mellin amplitude with the Borel transform. Inserting the Mellin pole expansion \cref{eq:decomMack} and performing the Borel transform \cref{eq:Borel} gives
\begin{equation} \label{eq:borelA}
    \mathcal{A}(S,T)=\frac{\lambda \Gamma(\Sigma-2)}{8} \lim_{\lambda\to \infty} \sum_{\mathcal{O}}  \int_{0}^{\infty} dx \, \frac{2\tau^2}{\sqrt{\lambda} S} \,  \frac{e^{\alpha_*}}{\alpha_*^{\Sigma-2}}  C_{\mathcal{O}} \, \mathcal{Q}^{\tau+2}_{\ell,x\tau^2}\left(\frac{\sqrt{\lambda}U}{2 \alpha_*} + \frac{2\Sigma-2}{3} -p_1-p_4 \right)\, ,
\end{equation}
where $\alpha_*$ is determined by the position of the pole in \cref{eq:decomMack} when we take the residue
\begin{equation}
    \alpha_*=\frac{\sqrt{\lambda } S}{2}\frac{1}{ \tau  +2 x \, \tau ^2 -\frac{2 \Sigma -2}{3}} \, ,
\end{equation}
and $U$ is a shifted flat-space Mandelstam variable such that $S+T+U = 0$.

Under this large twist approximation and Borel resummation, we arrive at the curvature expansion of the AdS Veneziano amplitude. The full AdS Veneziano amplitude then admits the following expansion,
\begin{equation}
    \mathcal{A}(S,T) =\sum_{k=0}^{\infty}\lambda^{- \frac{k}{2}} \mathcal{A}^{(k)} (S,T) \, ,
\end{equation}
where $\mathcal{A}^{(0)}$ corresponds to the flat-space Veneziano and higher $\mathcal{A}^{(k)}$ include $1/R^{2k}$ curvature corrections from large AdS radius expansion.

At each order in the curvature expansion, the amplitude $\mathcal{A}^{(k)}(S, T)$ exhibits poles at the physical location $S = \delta$, corresponding to string mass levels $m^2 \alpha' = \delta = 1,2,3,\cdots$. Near such a pole, the amplitude takes the generic form\footnote{The pole structure originates from the integral
\begin{equation*}
    \int_{0}^{\infty} dx \, e^{\frac{S}{4 \delta  x}-\frac{1}{4 x}} x^{-i} = \frac{(-4 \delta )^{i-1} \Gamma (i-1)}{(S-\delta)^{i-1}} \, ,
\end{equation*}
and the truncation at $3k+1$ arises naturally after performing the large-twist expansion.
}
\begin{equation}
\mathcal{A}^{(k)}(S, T) \sim  \sum_{i=1}^{3k+1}\frac{\mathcal{R}^{(k)}_i(\delta, T)}{(S - \delta)^{i}} + \mathcal{O}\left( (S - \delta)^0 \right),
\end{equation}
where the residue $\mathcal{R}^{(k)}_i(\delta, T)$ is a polynomial in $T$ encoding the finite spin support. We will explain the pole structure later. As an instance, the simplest flat-space Veneziano amplitude can be written as an infinite sum over the exchanged operators at each mass level $\delta$, 
\begin{equation}
\mathcal{A}^{(0)}(S, T) \sim  \frac{\mathcal{R}^{(0)}_1(\delta, T)}{S - \delta} \, ,\quad \mathcal{R}^{(0)}_{1}(\delta,T) =-\sum_{\ell=0}^{\delta-1}\frac{\langle f_0\rangle_{\delta, \ell} \,  C_{\ell }^{(1)}\left(\frac{2 T}{\delta }+1\right)}{\delta  (\ell +1)} \,.
\end{equation}
Here $\langle f_0\rangle_{\delta, \ell}$ is the leading-order OPE coefficient for the spin-$\ell$ operator at level $\delta$\footnote{Here we define averaged data $\langle X \rangle_{\delta,\ell}$ because the operators are degenerate, see discussion in \cref{app:OPEandpolynomial}.}, and $C^{(1)}_\ell(x)$ denotes the Gegenbauer polynomial arising from the Mack block. For comparison, the flat-space Veneziano amplitude itself is
\begin{equation}
    \mathcal{A}^{(0)}(S,T) = - \frac{\Gamma(-S) \Gamma(-T)}{ \Gamma(1-S-T)}  \, ,
\end{equation}
which exhibits a tower of simple poles at $S = \delta$ and $T = \delta$, and serves as the universal seed for all AdS amplitudes in the strong coupling limit.

We apply the method developed in \cite{Rigatos:2024beq,Wang:2024wcc} to evaluate the OPE coefficient $\langle f_0\rangle_{\delta ,\ell} $, where we use Harmonic numbers as a basis to describe the general stringy partial-wave coefficients. The result can be expressed as
\begin{equation}
    \langle f_0\rangle_{\delta ,\ell}=\sum_{k=0}^{\delta-1}\frac{ (-1)^{k+\ell } (k-\ell +1)_{\ell } \left(\frac{3}{2}+\ell \right)_{k-\ell }(\ell+1)}{4^{\ell } \Gamma(\ell+1) (3+2 \ell )_{k-\ell }}  {\delta^{k}}   Z_{k}(\delta-1) \, .
\end{equation}
Let us introduce the multiple harmonic numbers,
\begin{equation}\label{eq:hm}
    Z_{\{i_1,i_2,\cdots,i_k\}}(N) \equiv \sum_{n=1}^{N} \frac{ 1}{n^{i_1}}Z_{\{i_2,i_3,\cdots,i_k\}}(n-1)\,  ,
\end{equation}
where $i$ is called the letter and $(a)_b=\Gamma(a+b)/\Gamma(a)$ is called the Pochhammer symbol. We also define the special case that $Z_{\{\}}(N)=1$ and $Z_{\{i_1,\cdots,i_k\}}(0)=0$, $\forall\; k\geq 1$. Here we specialize to the case that every letter $i$ is $1$ and use the following shorthand
\begin{equation}
    Z_{k}(N)\equiv Z_{\{1_1,1_2,\cdots,1_k\}}(N)\; .
\end{equation}
This notation provides a concise representation for our purposes.

In the next section, we will combine the world-sheet using AdS$\times$S formalism to construct a finite ansatz of the first correction to the AdS Veneziano amplitude in the curvature expansion following the strategy pointed out by \cite{Wang:2025pjo}.

\section{World-sheet formalism and ansatz}\label{sec:worldsheet}

\subsection{World-sheet formalism}

The world-sheet and AdS$\times$S formalism of AdS amplitudes provides a powerful framework in which curvature corrections can be systematically bootstrapped. When external states carry Kaluza-Klein charge, their embedding into AdS$_5\times$S$^3$ modifies the structure of the world-sheet integrand. In this case, the amplitude can be expressed as a sum of integrals over single-valued multiple polylogarithms (SVMPLs), accompanied by rational prefactors determined by the flat-space Mandelstam variables $S/T/U$ and the R-symmetry Mellin variables $n_{ij}$.

Let us begin by recalling the representation of the curvature correction to the AdS Veneziano amplitude as an integral over the disk of the open string world-sheet \cite{Alday:2024yax,Alday:2024ksp},
\begin{equation}
    \mathcal{A}^{(k)}(S,T;\, n_{ij}) = \frac{1}{S+T} \int_{0}^{1} dz \, z^{-S-1} (1-z)^{-T-1} \, \mathcal{G}^{(k)}(S,T;\,z;\,n_{ij}) \, .
\end{equation}
Here, $S$ and $T$ are the flat-space Mandelstam variables, and $\mathcal{G}^{(k)}(S,T;\,z;\, n_{ij})$ is the integrand constructed from a finite basis of single-valued multiple polylogarithms (SVMPLs) evaluated along the real line $\zb=z$. Here we focus on the first curvature correction to the AdS Veneziano amplitude. The function $\mathcal{G}^{(1)}(S,T;\, z;\, n_{ij})$ is defined as
\begin{equation}\label{eq:ansatzG}
\mathcal{G}^{(1)}(S,T;\, z;\, n_{ij}) = \sum_{i=0}^{3}\sum_{\pm} \mathcal{F}^{\pm}_i(S,T;\, n_{ij}) \, \mathcal{T}^{\pm}_i(z).
\end{equation}
Each $\mathcal{T}^{\pm}_i(z)$ is built from SVMPLs with transcendental weight $i$, and $\mathcal{F}_i(S,T;\,n_{ij})$ is a rational function that encodes the internal symmetry information.

We choose the SVMPLs as our basis that eliminate monodromies around $z = 0$ and $z = 1$. The SVMPLs form a closed, finite-dimensional space at each weight. We consider the basis for describing the first curvature correction $\mathcal{A}^{(1)}(S,T;\, n_{ij})$ up to weight-3. To make the symmetry under $z\leftrightarrow 1-z$ manifest, we define the linear combinations,
\begin{equation}
\mathcal{L}^{\pm}_w(z) = \mathcal{L}_w(z) \pm \mathcal{L}_w(1-z) \, .
\end{equation}
On each transcendental weight $i$ we construct the basis following the conventions of \cite{Alday:2024yax,Alday:2024ksp} 
\begin{align}
    \mathcal{T}^{+}_{3}&= (\mathcal{L}^{+}_{000}(z),\, \mathcal{L}^{+}_{001}(z) , \,  \mathcal{L}^{+}_{010}(z) , \, \mathcal{L}^{+}_{011}(z) , \, \zeta(2)  \,\mathcal{L}^{+}_{0}(z) , \, \zeta(3)) \, , \nonumber \\
    \mathcal{T}^{-}_{3}&= (\mathcal{L}^{-}_{000}(z), \, \mathcal{L}^{-}_{001}(z) , \, \mathcal{L}^{-}_{010}(z) , \,  \mathcal{L}^{-}_{011}(z) , \, \zeta(2)    \,\mathcal{L}^{-}_{0}(z) ) \, , \mathcal{T}^{+}_{2}= (\mathcal{L}^{+}_{00}(z), \, \mathcal{L}^{+}_{01}(z) ,  \, \zeta(2)  ) , \, \nonumber \\
     \mathcal{T}^{-}_{2}&= (\mathcal{L}^{-}_{00}(z) ) , \quad
    \mathcal{T}^{+}_{1} = (\mathcal{L}^{+}_{0}(z)) , \quad \mathcal{T}^{-}_{1} = (\mathcal{L}^{-}_{0}(z)) , \quad \mathcal{T}^{+}_{0} = (1) , \quad \mathcal{T}^{-}_{0} = (0)  \,  ,
\end{align}
where we encode their explicit expressions\footnote{Our definition of the basis $\mathcal{L}^{\pm}_{w}(z)$ corresponds to the MPLs basis \cite{Alday:2024yax} with $L^{\pm}_{w}\to \mathcal{L}^{\pm}_{w}$. However, when expressed in terms of MPLs $L_{w}(z)$, the present functions differ from the $L^{\pm}_{w}(z)$ defined in \cite{Alday:2024yax} but also form a basis. Notice that $\mathcal{L}^{-}_{01}(z)=0$, indeed, in \cite{Alday:2024yax} we also don't need function basis $L^{-}_{01}(z)$. In the final result, we find that the coefficients of $\zeta_2$ are 0, which allows us to use a SVMPLs basis such as \cite{Alday:2024ksp}.} in the \cref{app:SVMPL}. Note that our basis includes functions with different transcendental weights, whose behavior is quite different from the known closed string amplitude.


\subsection{Construction of the World-sheet Ansatz}

With both the single-valued polylogarithm basis and the rational variables $(n_s, n_t, n_u, \Sigma)$ in place, we are now in a position to formulate the full ansatz for the first curvature correction to the AdS Veneziano amplitude. This ansatz reflects the combined spacetime and R-symmetry structure of the four-point correlator, and is constrained by both crossing symmetry and the analytic properties of the world-sheet integral. We assume that the world-sheet integrand can be written as \cref{eq:ansatzG}, where each $\mathcal{T}^{\pm}_{k}(z)$ is a single-valued multiple polylogarithm, chosen to span the symmetric and antisymmetric sectors under $z \leftrightarrow 1 - z$, and $\mathcal{F}_i$ are rational functions of Mandelstam variables and internal indices. Note that these transcendental functions encode the world-sheet monodromy structure. However, the integrated amplitude is no longer single-valued, because performing the $z$-integral breaks the world-sheet monodromy\footnote{One can already see this from flat-space Veneziano amplitude.}.

The rational functions $\mathcal{F}_k^{\pm}(S,T; n_{ij})$ are responsible for encoding the dependence on the Mandelstam variables $S$ and $T$, as well as the internal dependence $n_{ij}$ that arises from the R-symmetry structure on the sphere. Each rational term is constructed to be a polynomial in $S$ and $T$, divided by a universal denominator proportional to $S + T$. 

The structure of $\mathcal{F}_k^\pm$ is also constrained by the transcendental weight of the integrand. Each component of $L_k(z)$ contributes a fixed transcendental weight $k$; thus for the full amplitude to be homogeneous in weight (as expected for a fixed order in the curvature expansion), the accompanying $\mathcal{F}^{\pm}_k(S,T; n_{ij})$ must carry the remaining weight. This implies that the monomial $S^i T^{k-i}$ in $\mathcal{F}^{\pm}_k$ contributes weight $k$ from kinematics, to be matched with a corresponding $\zeta$-value in the low-energy expansion. Hence, the $S,T$ structure of $\mathcal{F}^{\pm}_k$ is tightly constrained by power counting arguments. More explicitly, we consider $\mathcal{F}^{\pm}_k$ that are constructed as
\begin{align}
\mathcal{F}^{\pm}_k = \frac{1}{S+T}\sum^{k}_{i=0} f^{\pm}_{i,k} S^{i} T^{k-i} \, ,
\end{align}
where the coefficients $f^{\pm}_{i,k}$ are linear functions of the internal variables 
\begin{equation}
    \{ n_s \, , n_t \, , n_u \, , \Sigma \, , \Sigma n_s \, , \Sigma n_t \, , \Sigma n_u\}.
\end{equation}
Notice that both $\mathcal{F}_k^\pm$ and $\mathcal{T}_k^\pm(z)$ are vectors of equal length. Imposing crossing symmetry leads to the constraints
\begin{align}
    \mathcal{F}^{+} (S,T; n_s,n_t) -\mathcal{F}^{+} (T,S; n_t,n_s) = 0\, , \\
    \mathcal{F}^{-} (S,T; n_s,n_t) + \mathcal{F}^{-} (T,S; n_t,n_s) =0 \, .
\end{align}

In the low-energy expansion $\lambda \to \infty$, this ansatz should reproduce the known super-gluon amplitude computed in \cite{Alday:2021odx} and the stringy corrections computed from the effective field theory method \cite{Glew:2023wik}. Indeed, our final result automatically reproduces these results.

In the next section, we will demonstrate how the unknown coefficients $f^{\pm}_{i,k}$ in this ansatz can be uniquely determined by matching the residues at physical poles $S = \delta$ with those derived from the superconformal block expansion. This matching serves as the core of our bootstrap strategy and completely determines the first curvature correction to the AdS Veneziano amplitude for arbitrary external KK modes.

\section{Bootstrap and exact result}\label{sec:exactres}

In this section, we derive the full first-order curvature correction to the AdS Veneziano amplitude for general Kaluza-Klein (KK) external modes. The derivation is based on a bootstrap strategy that matches the residues of S-channel poles in the conformal block expansion with those arising from the world-sheet ansatz constructed in \cref{sec:worldsheet}. This procedure uniquely determines the rational prefactors in the world-sheet integrand and thus fully fixes the amplitude.

The amplitude in Mellin space can be decomposed into superconformal blocks, whose Borel resummation yields the AdS amplitude in the flat-space limit. The poles in the Mellin variable $S$ correspond to the exchanged string states of mass level $\delta$, located at $S = \delta$. Near such a pole, the amplitude $\mathcal{A}^{(1)}(S,T)$ can be expanded as
\begin{equation}
    \mathcal{A}^{(1)}(S,T) \sim \sum_{i=1}^4 \frac{\mathcal{R}^{(1)}_{i}(\delta,T)}{(S-\delta)^{i}}+\mathcal{O}\left( (S-\delta)^0 \right)  \, ,
\end{equation}
where $\mathcal{R}^{(1)}_i(\delta, T)$ are polynomials in $T$ determined by the OPE data and anomalous dimensions of exchanged operators. The explicit expressions for these residues can be obtained from the Borel transform of the superblock decomposition, as described in \cref{app:OPEandpolynomial}. For example, the polynomial at the leading pole is
\begin{align}
    \mathcal{R}^{(1)}_{4}(\delta,T)=\sum_{\ell=0}^{\delta-1} \left(-\frac{4 \, \delta \, \langle f_{0} \rangle  _{\delta,\ell}}{\ell+1} \mathcal{C}^{0} (\delta,T) \right) = -\frac{4 \delta \,  (T+1)_{\delta -1}}{\Gamma (\delta )}\, .
\end{align}
where we have introduced a shorthand $\mathcal{C}^n(\delta,T)$,
\begin{equation}
\mathcal{C}^n(\delta,T)\;\equiv\;\left(\frac{d}{dT}\right)^{ n}\,
C^{(1)}_{\ell} \left(1+\frac{2T}{\delta}\right)\, .
\end{equation}
The triple pole reads
\begin{equation}
    \mathcal{R}^{(1)}_{3}(\delta,T)=\sum_{\ell=0}^{\delta-1} \left(-\frac{4 \delta  \langle f_0\rangle_{\delta ,\ell}}{2 (\ell +1)} \mathcal{C}^{1}  (\delta,T)  +\frac{4 (\Sigma -4) \langle f_0\rangle_{\delta ,\ell} }{3 (\ell +1)} \mathcal{C}^{0}  (\delta,T) \right) \, .
\end{equation}
The general result for the residue is provided in the \cref{app:OPEandpolynomial}.

From the world-sheet perspective, the pole structure arises from the behavior of the integrand near the OPE limit $z \to 0$. In this limit, the expansion of SVMPLs such as $\log^k z$ dominates, producing a sequence of higher-order poles. A simple example illustrates the origin of the $k+1$-th order pole,
\begin{equation}
    \int_0^1 dz \, z^{-S+\delta-1}  \log^k(z) = - \frac{\Gamma (k+1)}{(S-\delta)^{k+1}} .
\end{equation}
This shows that logarithmic singularities in the world-sheet integrand directly reproduce the expected pole structure in the AdS Veneziano amplitude.

To fix the unknown coefficients $\mathcal{F}_k^\pm(S, T; n_{ij})$ in the world-sheet ansatz, we expand the integrand near $z = 0$ and extract the polynomials over poles at $S = \delta$. By matching the residues computed from the superconformal block expansion and those poles arising from the integrals of SVMPLs, we determine the coefficients in the rational prefactors of each SVMPL. This matching uniquely fixes the entire ansatz.

The coefficient $\mathcal{F}^{\pm}_{i}$ can be simplified by introducing 
\begin{equation}
    N_{s} = n_s -\frac{\Sigma-3}{3} \, , \quad N_{t} = n_t -\frac{\Sigma-3}{3} \, , \quad N_{u} = n_u -\frac{\Sigma-3}{3} \, , 
\end{equation}
and the final expressions are given by
\begin{align}
    \mathcal{F}^{+}_3&= \left(\frac{1}{4} \left(S^2+T^2\right),0,-\frac{1}{4} (S+T)^2,\frac{1}{4} \left(-S^2-T^2\right),0,2 \left(S^2+S T+T^2\right) \right) \, ,\\
    \mathcal{F}^{-}_3&= \left(\frac{1}{4} \left(S^2-T^2\right),0,\frac{1}{4} \left(T^2-S^2\right),\frac{1}{4} \left(S^2-T^2\right),0 \right) \, ,\\
    \mathcal{F}^{+}_2&= \left(-\frac{S^2+8 S T+T^2}{12 (S+T)}-\frac{1}{6} \Sigma  (S+T),\frac{1}{3} \Sigma  (S+T)-\frac{S^2-4 S T+T^2}{12 (S+T)},0 \right) \, , \\
    \mathcal{F}^{-}_2&=\left(-\frac{1}{12} (2 \Sigma +1) (S-T) \right) \, , \quad \mathcal{F}^{+}_1 = \frac{1}{12} \left(5-6 (\Sigma -3) (N_s+N_t)\right) \, , \\
    \mathcal{F}^{-}_1&=- \frac{ (\Sigma -3) \left(N_s-N_t\right)}{2 } -\frac{(4 \Sigma -1) (S-T)}{12 (S+T)} \, ,  \mathcal{F}^{+}_0 =\frac{\frac{5}{3}-2 (\Sigma -3) \left(N_s+N_t\right)}{S+T}  \, .
\end{align}
This result not only reduces to the known $\langle \mathcal{O}_{2}\mathcal{O}_{2}\mathcal{O}_{2}\mathcal{O}_{2}\rangle$ case \cite{Alday:2024yax} where $n_s=n_t=n_u=0$ and $\Sigma=4$, but also extends it to arbitrary external KK configurations. Furthermore, the rational structure of $\mathcal{F}^{\pm}_k$ and the emergence of SVMPLs reinforce the deep connection between world-sheet geometry and CFT bootstrap data. 

\begin{align}
    \langle f_0 \tau_2 \rangle_{\delta,\delta-1} =&\frac{r_{0}(\delta)}{4 \sqrt{\delta}} \left(-4 n_s (\Sigma -3)+3 \delta^2-3 \delta-I_p-12 \Sigma +2 J_p +18\right) \, ,  \\
    \langle f_2 \rangle_{\delta,\delta-1} 
    =& \frac{r_0(\delta)}{\delta}\left[ -\frac{7 \delta ^3}{12}  +\delta  \left(I_p \left(\Sigma -\frac{5}{2}\right)+\Sigma  (11-4 \Sigma )-\frac{25}{6} \right) \right. \nonumber\\
    & \qquad -\frac{1}{6} \Sigma  (3 I_p+6 J_p+58)+\frac{1}{8} I_p (I_p+6)+2 J_p+\frac{2 \Sigma ^3}{3}+2 \Sigma ^2+\frac{49}{8} \nonumber \\
    & \left.  \qquad +\delta ^2 \left(2 (\Sigma -4) \Sigma +\frac{33}{4}\right)  + \delta n_s (6-2 \Sigma ) \right]  + \delta^2 \zeta(3) \langle f_0 \rangle_{\delta,\delta-1} \, ,
\end{align}
where $I_p=(p_1-p_2)^2+(p_3-p_4)^2$, $J_p=p_1^2+p_2^2+p_3^2+p_4^2$ and
\begin{equation}
    r_{n}(\delta)=\frac{4^{1-\delta } \delta ^{\delta -2 n-1} (\delta -2 n)^2}{\Gamma (-n+\delta +1)} \, .
\end{equation}
Here $\langle f_0 \tau_2\rangle_{\delta,\ell}$ and $\langle f_2\rangle_{\delta,\ell}$ denote averaged OPE data for an exchanged operator at level $\delta$ and spin $\ell$. Crucially, \cref{eq:R4,eq:R3,eq:R2,eq:R1} are sufficient to fix $\mathcal{A}^{(1)}$ from the given $\langle f_0\rangle_{\delta,\ell}$. The remaining averaged data $\langle f_0\tau_2\rangle_{\delta,\ell}$ and $\langle f_2\rangle_{\delta,\ell}$ are not independent inputs but are solved from residues matching. For the leading trajectory $\ell=\delta-1$, we have given explicit results. More related discussions can be found in \cref{app:OPEandpolynomial}.

Our method is fully constructive: starting from the ansatz built from symmetry principles and SVMPL basis, the entire amplitude is uniquely fixed by matching a finite number of residues. This method can, in principle, be extended to higher curvature orders, to closed-string scattering, and to other backgrounds.

\section{Low- and high-energy limit}\label{sec:limit}

\subsection{Low-energy limit}
In this section, we analyze the low-energy behavior of the AdS Veneziano amplitude by expanding the world-sheet integral around the point $S = T = 0$. The expansion is organized as a series in the large 't Hooft coupling $\lambda^{-1/2}$, where each term corresponds to a higher-derivative correction in the dual AdS effective action.

To extract this expansion systematically, we begin with the 1d disk integral of the form
\begin{equation}
I_w(S,T) = \int_{0}^{1} d z\, z^{-S-1} (1 - z)^{-T-1} \, L_w(z),
\end{equation}
where each integrand ${L}_w(z)$ is a multiple polylogarithm. The dependence on the Mandelstam variables $S$ and $T$ enters through the Koba-Nielsen factor. The general evaluation of such integrals at small $S$ and $T$ was presented in \cite{Alday:2024yax}, and takes the form
\begin{align}
    I_w  = \text{pole} +
    \sum_{i,j=0} (-S)^i (-T)^j \sum_{W\in 0^i \shuffle 1^j \shuffle w}  \left( {L}_{0W}(1) - {L}_{1W}(1) \right) ,
\end{align}
where $\shuffle$ denotes the shuffle product and pole represents the polar contribution. One can also explicitly verify this expansion by directly computing the 1d integral. For the correlator we consider, the relevant functions include $\log^kz$, $\log^k(1-z)$, and $\operatorname{Li}_k(z)$. The first two contributions can be obtained by acting with $(-\partial_S)^k$ and $(-\partial_T)^k$ on the basic integral,
\begin{equation}
    I_{\{\varnothing\}}= \frac{\Gamma (-S) \Gamma (-T)}{\Gamma (-S-T)} = -\frac{1}{S}-\frac{1}{T}+ \zeta(2) \, (S+T) +2 \zeta (3) \, S T   + \cdots \, ,
\end{equation}
where $\zeta(n)$ denotes the zeta value. For integrals involving $\operatorname{Li}_k(z)$, we employ a generating-function trick. For example, for $\operatorname{Li}_2(z)$ we have
\begin{align}
    I_{01}  &= -\int_{0}^{1}dz \, z^{-S-1}(1-z)^{-T-1} \text{Li}_{2}(z) = - \sum_{k=1}^{\infty}\int_{0}^{1}dz \, z^{-S-1}(1-z)^{-T-1} \frac{z^k}{k^2} \nonumber \\
    &= - \sum_{k=1}^{\infty}  \frac{\Gamma (-T) \Gamma (k-S)}{k^2 \Gamma (k-S-T)} = \frac{\zeta(2)}{T}+  \zeta(3) +\frac{7}{4} \zeta(4) T -\frac{7}{4} \zeta(4) S + \cdots
\end{align}

By explicitly expanding the integrated world-sheet correlator, we derive a systematic low-energy series in large 't Hooft coupling $\lambda^{-1/2}$. The first few terms of the amplitude $\mathcal{A}^{(1)}(S,T)$ are given by
\begin{align}\label{eq:lowenergyA}
    \mathcal{A}^{(1)}(S,T)=&\frac{2 (\Sigma -3) (3 n_s -\Sigma +4)}{3 S^2 T}+\frac{2 (\Sigma -3) (3 n_t -\Sigma +4)}{3 S T^2}  \nonumber \\
    &+ \frac{2}{3} (\Sigma -3) (3 n_s +3 n_t-2 \Sigma +8) \, \zeta(3) \nonumber
    \\
    &+ \left(\frac{1}{2} (8 n_s (\Sigma -3)+n_t(\Sigma -3)+\Sigma  (19-3 \Sigma )-43) S \right. \\  
    & \qquad \left. +\frac{1}{2} (n_s (\Sigma -3)+8 n_t (\Sigma -3)+\Sigma  (19-3 \Sigma )-43) T \right) \zeta(4) \nonumber \\
    &+ \cdots \nonumber
\end{align}
This expansion corresponds to a Taylor series in $S, T \ll 1$, valid in the near flat-space limit of the AdS curvature correction.

Before proceeding, we revisit the low-energy behavior of the Mellin amplitude $\mathcal{M}$, following the methodology developed in \cite{Wang:2025pjo}. As in the case of $\mathcal{N}=4$ SYM, the Borel transform of $\mathcal{M}$ reveals that the dominant contributions arise from specific powers of $s$, $t$ and $u$. In particular, the large-$p$ limit \cite{Aprile:2020luw} imposes an upper bound on the degree of $\bs$, while perturbative data impose a lower bound. A similar thing happens here. These two constraints ensure that only a small part of $\mathcal{M}$ contributes to $\mathcal{A}^{(1)}$, allowing us to organize terms according to their $\bs/\bt/\bu$ degree at each order in $\lambda^{-1/2}$. At order $\lambda^{-\frac{i}{2}}$, the leading contributions to $\mathcal{M}^{(i)}$ are homogeneous polynomials in $\bs$, $\bt$, and $\bu$ of degree $i-2$, and are fixed by the flat-space amplitude $\mathcal{A}^{(0)}$. Subleading terms involve mixed invariants of lower degree, symmetric under $\bs \leftrightarrow \bt$. The method to construct a complete basis of symmetric invariants is provided in \cite{Aprile:2020mus}.

In terms of the canonical variables $\tilde{s}=(n_s-\Sigma)/2$ and $\tilde{t}=(n_t-\Sigma)/2$, the amplitude $\mathcal{M}(s,t)$ admits the following expansion \cite{Glew:2023wik}
\begin{align}\label{eq:lowenergyM}
    \mathcal{M}(s,t) =& -\frac{1}{2} \frac{1}{(\bs+1) (\bt+1)} + \lambda^{-1} \, 2^3 \,(\Sigma-2)_2 \, \zeta(2) \nonumber \\
    &+ \lambda^{-\frac{3}{2}} \, 2^{5}  \, \left((\Sigma-2)_3( \bs+\bt )-3(\Sigma-2)_2( \tils+\tilt)+ a _1(\Sigma-2)_2 \right)  \, \zeta(3) \nonumber \\
    &+ \lambda^{-2} \, 2^{7} \, \zeta(4) \left[ (\Sigma-2)_4 \left( \frac{7}{8} \bs^2 + \frac{7}{8}\bt^2+\frac{1}{8} \bu^2 \right) \right. \nonumber \\
    &\qquad \left. + (\Sigma -2)_3 \left( b_1(\bs+\bt-3 (\tils+\tilt))+e_1(\bs+\bt)+f_1\bu \right) +\cdots \right] +\mathcal{O}(\lambda^{-\frac{5}{2}}) \, .
\end{align}
where $a_1$, $b_1$, $e_1$ and $f_1$ denote the unknown coefficients, and the omitted terms at $\lambda^{-2}$ include lower-degree polynomials in $\bs, \bt$ that are controlled by higher curvature corrections. In principle, these Wilson coefficients\footnote{See the Wilson coefficients defined in \cite{Glew:2023wik}, which are the coefficients before higher-derivative effective interactions.} correspond to higher-derivative couplings in an eight-dimensional effective action, and thus they determine the coefficients in the low-energy expansion of the amplitude; however, that approach merely provides several constraints between Wilson coefficients.

By contrast, our bootstrap-based world-sheet approach yields a unique solution for sub-leading Wilson coefficients. After applying Borel transform \cref{eq:Borel}, the first curvature correction derived here not only reproduces the super-gluon contribution \cref{eq:lowenergyM}, but also provides concrete predictions for the sub-leading terms. By comparing \cref{eq:lowenergyM} with \cref{eq:lowenergyA}, we find that
\begin{equation}
    a_1=-4 \, , \qquad \qquad  f_1=b_1+e_1+\frac{21}{8} \, .
\end{equation}
Specifically, we determine the $\lambda^{-\frac{3}{2}}$ term and all sub-leading Wilson coefficients in the correlator studied. Our predictions match the localization method in $\langle \mathcal{O}_{2} \mathcal{O}_{2} \mathcal{O}_{2} \mathcal{O}_{2}\rangle$ case \cite{Behan:2023fqq}. These predictions also serve as nontrivial checks for future localization computations and offer a systematic path toward higher-order curvature corrections.

\subsection{High-energy limit}

We now turn to the high-energy behavior of the AdS stringy amplitude, where the Mandelstam variables $S$ and $T$ become large with fixed ratio $S/T$. Remarkably, we find that the leading high-energy behavior of the AdS Veneziano amplitude is \textit{universal} for all Kaluza–Klein configurations we studied. In particular, its high-energy limit matches exactly with the $\langle\mathcal{O}_2\mathcal{O}_2\mathcal{O}_2\mathcal{O}_2\rangle$ configuration \cite{Alday:2024yax}. The same universality also holds for the AdS Virasoro-Shapiro amplitude, which governs closed string scattering in the AdS$_5\times$S$^5$ background \cite{Alday:2023pzu}.

This universality implies a simple and elegant relation between the open and closed string high-energy limits,
\begin{equation}
    \mathcal{E}_{\text{open}}(S,T) = \frac{1}{2} \mathcal{E}_{\text{closed}}(4S,4T),
\end{equation}
where $\mathcal{E}_{\text{open}}(S,T)$ and $\mathcal{E}_{\text{closed}}(S,T)$ denote the high-energy exponents controlling the asymptotic behavior of the open and closed AdS amplitudes, respectively. Explicitly, using a saddle-point approximation for the $z$-integral with saddle point at $z_0 = \frac{S}{S+T}$, one finds that the exponent controlling the amplitude takes the form given by
\begin{align}
    A^{\text{high}}\sim \exp\left( -\mathcal{E}_{\text{open/closed}}(S,T)\right) = \exp\left(- \mathcal{E}^{(0)}(S,T) - \frac{S^2}{R^2} \mathcal{E}^{(1)}(S,T) - \mathcal{O}\left(\frac{1}{R^4}\right)\right)  \, .
\end{align}
For the open string scattering, we have
\begin{align}
    \mathcal{E}^{(0)}(S,T)_{\text{open}}=&  S \log(S)+T \log(T) -(S+T) \log(S+T) \, , \\
    \mathcal{E}^{(1)}(S,T)_{\text{open}}=& \frac{1}{2}  \mathcal{L}_{000}(z_0) - \frac{1}{2}  \mathcal{L}_{001}(z_0) - \frac{z_0}{2}  \mathcal{L}_{010}(z_0) - \frac{(z_0 -1)^2}{2 z_0^2}  \mathcal{L}_{011}(z_0) \nonumber \\
    &+ \frac{z_0-1}{2 z_0^2}  \mathcal{L}_{101}(z_0)  + \frac{(z_0-1)^2}{2 z_0^2}  \mathcal{L}_{111}(z_0) + \zeta(3) \, .
\end{align}
These expressions match the behavior derived from the flat-space Veneziano and Virasoro-Shapiro amplitudes, and their appearance here confirms that the universality of AdS string amplitude maintains in the high-energy regime, independent of the external KK structure. The equality $ \mathcal{E}_{\text{open}} = \frac{1}{2} \mathcal{E}_{\text{closed}} $ is therefore a nontrivial prediction of the AdS string framework, robust in general backgrounds  \cite{Chester:2024esn,Wang:2025pjo,Jiang:2025oar}.

\section{Discussion}

This work provides a convenient bootstrap method of the AdS Veneziano amplitude with \textit{arbitrary} external KK charges at the first nontrivial order in the curvature expansion. The conceptual hinge is the structural equivalence we established between the superconformal block decomposition and a world-sheet ansatz in AdS$\times$S space, once the Borel transform is taken in the large-twist regime. In practice, matching the Borel-transformed superblock residues to the OPE-limit logarithms of integrand rigidly fixes the unknown coefficients and yields a \textit{unique} solution for $\mathcal{A}^{(1)}$ without any low-lying KK input. The method reduces the initial problem to a small and overconstrained linear problem.

We discuss the low-energy expansion and fix the entire infinite tower of sub-leading Wilson coefficients. At the opposite side, in the high-energy limit, the saddle analysis of the disk integral shows a universal exponent that is independent of the external KK charges and is related to its closed-string counterpart by $\mathcal{E}_{\text{open}}(S,T)=\tfrac12 \mathcal{E}_{\text{closed}}(4S,4T)$, reinforcing the idea that AdS world-sheet dynamics retains a flat-space-like organizing principle even in the presence of AdS curvature.

Methodologically, the main advance in our approach is that the KK dependence can be handled \textit{uniformly} through the AdS$\times$S Mellin variables, with single-valuedness provided by the SVMPL basis. The resulting uniqueness of $\mathcal{A}^{(1)}$ across all KK configurations is, in our view, a strong indication that the pipeline
\[
\text{superblocks} \;\xRightarrow[\text{Borel}]{\text{large twist}}\; \text{AdS}\times\text{S  world-sheet}
\;\Longrightarrow\; \text{unique first curvature correction}
\]
is not an accident of the specific background, but a structural feature of supersymmetric holographic theories in the small curvature regime. It is natural to study whether the same mechanism operates in more backgrounds.

There are several directions for future work. A first goal is to elevate the present analysis to the next order in the curvature expansion. 
Technically, this requires enlarging the SVMPL basis up to weight six; conceptually, it tests whether the AdS stringy structure of the world-sheet ansatz persists beyond $\mathcal{A}^{(1)}$.

A second avenue is to apply the same bootstrap directly to the closed-string sector with KK charges, and to search for an \emph{AdS KLT} relation. Related aspects from a function-basis perspective have been explored in \cite{Alday:2025bjp,Baune:2025hfu}.

It is natural to push the present analysis beyond the planar limit. Loop-level super-gluon amplitudes are now available \cite{Alday:2021ajh,Huang:2023oxf,Behan:2023fqq,Huang:2023ppy,Huang:2024rxr}, and mixed gluon–graviton results have also appeared \cite{Chester:2025ssu}. One would expect that genus-one calculations involve a richer functional alphabet, making any world-sheet ansatz and its bootstrap matching more intricate. 

Our present work focuses on the four-point case, where superconformal Ward identities efficiently constrain the kinematics and R-symmetry structures, allowing a finite single-valued basis and a uniquely fixed $\mathcal{A}^{(1)}$. A sharp next step is to investigate higher-point stringy corrections \cite{VilasBoas:2025vvw} and test whether the same AdS$\times$S organization using world-sheet building blocks continues to constrain the bootstrap.

On the CFT side, our predictions for the low-energy Wilson coefficients are consistent with results from localization and integrability. They can also be tested against bootstrap computations for correlators with higher KK charges. The residue formulas provide direct access to averaged OPE data; it would be interesting to disentangle individual multiplet contributions with a refined analysis of the spherical variables and string spectrum.

\vspace{0.8em} 
    
	\noindent{\bf Acknowledgments.}
	    The author would like to thank Shai Chester, Xiaoyu Fa, Tobias Hansen, Zhongjie Huang, Maria Nocchi and Deliang Zhong for useful discussions, and Ellis Ye Yuan for valuable suggestions on the first draft. We thank Xiaoyu Fa and Ellis Ye Yuan for collaborations on related projects. BW is supported by the National Science Foundation of China under Grant No.~124B2095, No.~12175197 and Grant No.~12347103.

\appendix

\section{Mack polynomials}
We will provide the explicit expression of Mack polynomials in Mellin representation in this appendix. The superblock decomposition of the Mellin amplitude can be written as
\begin{equation} \label{eq:decomMackapp}
     \mathcal{M}(s,t) \sim \sum_{\mathcal{O}} \sum_{m=0}^{\infty}   \frac{ C_{\mathcal{O}} }{s - \tau - 2m } \, \mathcal{Q}^{\tau+2}_{\ell,m}(u-p_1-p_4) \, ,
\end{equation}
where $\mathcal{Q}^{\tau+2}_{\ell,m}(u)$ can be written as
\begin{equation}
    \mathcal{Q}^{\tau+2}_{\ell,m}(u) = \kappa^{p_1+1,p_2+1,p_3+1,p_4+1}_{\ell,m,\tau+2}  Q^{p_2-p_1,p_3-p_4,\tau+2}_{\ell,m}(u) \, .
\end{equation}
We define the Mack polynomials as follows \cite{Dolan:2011dv,Mack:2009mi,Dey:2017fab},
\begin{equation}
    Q^{\Delta_{21},\Delta_{34},\tau}_{\ell,m}(s)=\frac{2^{\ell } \ell ! \Gamma (\ell +\tau -1)}{{\Gamma (2 \ell +\tau -1)}}\sum _{k=0}^{\ell } \sum _{n=0}^{\ell -k} (-m)_k \left(m+\frac{s+\tau }{2}\right)_n \mu^{\Delta_{21},\Delta_{34}} (\ell ,k,n,\tau ) \, ,
\end{equation}
where the function $\mu^{\Delta_{21},\Delta_{34}} (\ell ,k,n,\tau )$ is defined as
\begin{align}
    \mu^{\Delta_{21},\Delta_{34}} (\ell ,k,n,\tau ) = & (-1)^{k+n+\ell } \frac{ (\ell +\tau -1)_n}{k! n! (-k-n+\ell )!}  \left(\frac{\Delta_{21}+\Delta_{34}}{2}+n+1\right)_k \nonumber  \\
    & \times \left(\frac{\Delta_{21}}{2}+k+n+\frac{\tau }{2}\right)_{-k-n+\ell }  \left(\frac{\Delta_{34}}{2}+k+n+\frac{\tau }{2}\right)_{-k-n+\ell } \, \nonumber \\
    & \times _4F_3\left(\begin{array}{c}
        -k,-n-\ell -1,-\frac{\Delta_{21}}{2}+\frac{\tau }{2}-1,-\frac{\Delta_{34}}{2}+\frac{\tau }{2}-1  \\
        -\ell ,-\frac{\Delta_{21}}{2}-\frac{\Delta_{34}}{2}-k-n,\tau -2
    \end{array};1\right) \, .
\end{align}
Finally, the normalization factor $\kappa$ reads
\begin{align}
    \kappa^{p_1,p_2,p_3,p_4}_{\ell,m,\tau} & =-\frac{2^{1-\ell } (\ell +\tau -1)_{\ell } \Gamma (2 \ell +\tau )}{m! (\ell +\tau -1)_m } \frac{1}{\Gamma \left(-m+\frac{p_1}{2}+\frac{p_2}{2}-\frac{\tau }{2}\right) \Gamma \left(-m+\frac{p_3}{2}+\frac{p_4}{2}-\frac{\tau }{2}\right)} \nonumber \\
    &\times \frac{1}{ \Gamma \left(\frac{p_2-p_1}{2}+\ell +\frac{\tau }{2}\right) \Gamma \left(\frac{p_1-p_2}{2}+\ell +\frac{\tau }{2}\right) \Gamma \left(\frac{p_4-p_3}{2}+\ell +\frac{\tau }{2}\right) \Gamma \left(\frac{p_3-p_4}{2}+\ell +\frac{\tau }{2}\right)} \, .
\end{align}
\section{Single-valued multiple polylogarithms}\label{app:SVMPL}

Multiple polylogarithms (MPLs) $L_w(z)$ form a class of iterated integrals labeled by a word $w = (a_1, a_2, \dots, a_n)$ consisting of ``letters'' $a_i \in \mathbb{C}$. MPLs can be recursively defined via iterated integrals \cite{Goncharov:1998kja,Goncharov:2001iea}
\begin{align*}
	L_{a_1a_2\ldots a_n}(z)=\int_0^z\frac{\mathrm{d}t}{t-a_1}L_{a_2a_3\ldots a_n}(t)\,, 
\end{align*}
with $L_\varnothing(z) = 1$. The length $n$ of the word $w$ defines the \textit{transcendental weight} of the MPL. 

In particular, if we restrict ourselves to words composed only of letters $\{0,1\}$, which is sufficient for describing open-string amplitudes with boundary insertions at $z=0,1,\infty$. In this alphabet, the MPLs satisfy certain differential relation
\begin{equation*}
    \partial_z L_{0w}(z) = \frac{1}{z} L_w(z)\,, \quad \partial_z L_{1w}(z) = \frac{1}{z-1} L_w(z)\, .
\end{equation*}
Let us give some special cases,
\begin{align*}
	L_{\varnothing}(z)=1\,,\qquad
		    L_{0_n}(z)=\frac{1}{n!}\log^nz\,.
\end{align*}

In general, MPLs have nontrivial monodromy around their singularities at $z=0,1,\infty$, which is incompatible with single-valuedness on the punctured Riemann sphere. To obtain well-defined string integrands, we instead work with \textit{single-valued multiple polylogarithms} (SVMPLs), denoted $\mathcal{L}_w(z)$, which are real-analytic and free of branch cuts.

These functions can be constructed via a single-valued map $\mathrm{sv}$ acting on MPLs, as developed in \cite{Brown:2004ugm}. The resulting $\mathcal{L}_w(z)$ are linear combinations of products $L_{w'}(z)$ and $L_{w''}(\bar z)$ that preserve both the weight and the algebraic structure of the original MPL
\begin{equation}
    \mathcal{L}_w(z) = \mathrm{sv}[L_w(z)] .
\end{equation}
In the context of AdS open string scattering, the world-sheet integrals are evaluated over the real interval $z \in [0,1]$. Consequently, the integration domain is constrained to lie along the real axis, and we always evaluate SVMPLs at $\bar z = z \in [0,1]$. Hence, the SVMPLs reduce to real-valued functions of a single real variable. This makes them ideal building blocks for constructing open string world-sheet integrands.

The computation and manipulation of SVMPLs can be efficiently handled using the \texttt{PolyLogTools} package \cite{Duhr:2019tlz}. In this package, SVMPLs are accessed using the command \texttt{cG}, and we define
\begin{equation}
    \mathcal{L}_{a_1 a_2 \cdots a_n}(z) \equiv \texttt{cG[a\_1,a\_2,\dots,a\_n,z]}.
\end{equation}
We use a fixed basis up to weight-3 SVMPLs with $a_i \in \{0,1\}$, organized by parity under $z \leftrightarrow 1-z$ into symmetric and antisymmetric sectors,
\begin{align}
    \mathcal{L}^{\pm}_w(z) = \mathcal{L}_w(z) \pm \mathcal{L}_w(1-z) \, .
\end{align}
The specific basis elements used in our bootstrap analysis are listed explicitly as (ignore z)
\begin{align}
    \mathcal{L}^{\pm}_{000}(z)&= 8 L_{000} \pm  8L_{111} \, , \\ 
    \mathcal{L}^{\pm}_{001}(z)&=2L_{001} +2L_{010}\pm4L_{011}+4L_{100}\pm2L_{101}\pm2L_{110}\mp2\zeta(3) \, , \\
    \mathcal{L}^{\pm}_{010}(z)&=4L_{001}+4L_{010}\pm4L_{101}\pm4L_{110}\pm4\zeta(3)\, ,\\
    \mathcal{L}^{\pm}_{011}(z)&=\pm2 L_{001}\pm2 L_{010}+4 L_{011}\pm4 L_{100}+2 L_{101}+2 L_{110}\pm2 \zeta (3) \, ,\\
    \mathcal{L}^{\pm}_{00}(z)&= 4 L_{00}\pm4 L_{11} \, , \quad 
    \mathcal{L}^{+}_{01}(z)= 4 L_{01}+4 L_{10} \, , \quad \mathcal{L}^{\pm}_{0}(z)= 2 L_{0}\pm 2 L_{1} \, .
\end{align}
These functions serve as the fundamental transcendental building blocks of the curvature-corrected AdS Veneziano amplitude.

\section{OPE data and polynomial residues}\label{app:OPEandpolynomial}

At fixed string level $\delta$ and spin $\ell$, the exchanged operators are highly degenerate. They differ by internal quantum numbers and by short multiplet recombinations; yet, they enter the four-point function only through a finite set of \textit{averaged} combinations. It is therefore convenient to introduce the averaged data $\langle X \rangle_{\delta,\ell}$.

We expand the twist and OPE coefficients as in \cref{eq:borelA} from \cref{eq:largetau1,eq:largetau2,eq:largetau3}. Imposing that the contribution at order $\lambda^{-1/4}$ to the amplitude vanishes,
\begin{equation}
\mathcal{A}^{(1/2)}(S,T)=0,
\end{equation}
one finds two universal constraints on the averaged OPE data
\begin{equation}
\langle \tau_1 \rangle_{\delta,\ell} = -\,\ell \;,\qquad
\langle f_1 \rangle _{\delta,\ell} = \frac{ 4( \Sigma -2) \ell +(p_1-p_2)^2+(p_3-p_4)^2 -1}{2 \sqrt{\delta}}\langle f_0 \rangle_{\delta,\ell} \; .
\label{eq:C1-leading}
\end{equation}
All unknown OPE data first appear at order $\lambda^{-1}$ through $\mathcal{A}^{(1)}$. Near $S=\delta$, the order-$\lambda^{-1}$ amplitude $\mathcal{A}^{(1)}(S,T)$ has the generic expansion
\begin{equation}
\mathcal{A}^{(1)}(S,T) \;\sim\;
\sum_{i=1}^{4}\;\frac{\mathcal{R}^{(1)}_{i}(\delta,T)}{(S-\delta)^{i}}
\;+\; \text{regular}\,,
\end{equation}
where each $\mathcal{R}^{(1)}_{i}$ is a polynomial in $T$ with finite spin support. Introducing the shorthand
\begin{equation}
\mathcal{C}^n(\delta,T)\;\equiv\;\left(\frac{d}{dT}\right)^{ n}\,
C^{(1)}_{\ell} \left(1+\frac{2T}{\delta}\right)\,,
\end{equation}
one finds the following expressions for the quartic and cubic poles in the main text,
\begin{align} \label{eq:R4}
    \mathcal{R}^{(1)}_{4}(\delta,T) & =\sum_{\ell=0}^{\delta-1} \left(-\frac{4 \, \delta \, \langle f_{0} \rangle  _{\delta,\ell}}{\ell+1} \mathcal{C}^{0} (\delta,T) \right) \, , \\
    \mathcal{R}^{(1)}_{3}(\delta,T) & =\sum_{\ell=0}^{\delta-1} \left(-\frac{4 \delta  }{2 (\ell +1)} \langle f_0\rangle_{\delta ,\ell} \, \mathcal{C}^{1}  (\delta,T)  +\frac{4 (\Sigma -4) }{3 (\ell +1)} \langle f_0\rangle_{\delta ,\ell}\, \mathcal{C}^{0}  (\delta,T) \right) \, . \label{eq:R3}
\end{align}
The quadratic pole can be written in a compact closed form that makes its dependence on the averaged OPE data manifest,
\begin{align} \label{eq:R2}
    \mathcal{R}^{(1)}_{2}(\delta,T)  &= \sum_{\ell=0}^{\delta-1} \left( \frac{T^2+\delta  T }{\delta  (\ell +1)} \langle f_0\rangle_{\delta ,\ell} \, \mathcal{C}^2(\delta,T) + \frac{8 \delta  \Sigma -17 \delta +8 \Sigma  T-2 T}{6 \delta  (\ell +1)} \langle f_0\rangle_{\delta ,\ell} \,  \mathcal{C}^1(\delta,T) \right.  \nonumber \\
    &  \left. \quad -\frac{3 I_p +4 \Sigma ^2-6 J_p +8}{6 \delta  (\ell+1)}  \langle f_0\rangle_{\delta ,\ell} \,  \mathcal{C}^0(\delta,T)  -\frac{2 \langle f_0 \tau_2 \rangle_{\delta ,\ell} }{\sqrt{\delta } (\ell +1)} \mathcal{C}^0(\delta,T) \right) . 
\end{align}
The simple pole reads
\begin{align} \label{eq:R1}
    \mathcal{R}^{(1)}_{1}(\delta,T)  =&   \sum_{\ell=0}^{\delta-1} \left( -\frac{T^3}{3 \delta ^2 (\ell +1)}  \langle f_0\rangle_{\delta ,\ell} \,  \mathcal{C}^3(\delta,T) \right.   +\frac{P_2}{6 \delta ^2 (\ell +1)}  \langle f_0\rangle_{\delta ,\ell} \,  \mathcal{C}^2(\delta,T) \nonumber  \\
    &+\frac{P_1}{\delta ^2 (\ell +1)} \langle f_0\rangle_{\delta ,\ell} \,  \mathcal{C}^1(\delta,T) +\frac{P_0}{\delta ^2 (\ell +1)} \langle f_0\rangle_{\delta ,\ell} \,  \mathcal{C}^0(\delta,T) - \frac{\langle f_2\rangle_{\delta ,\ell}}{\delta  (\ell +1)} \mathcal{C}^0(\delta,T) \nonumber \\
    & \left. +\frac{2 T}{\delta ^{3/2} (\ell +1)} \langle f_0 \tau_2\rangle_{\delta ,\ell} \,  \mathcal{C}^1(\delta,T) + \frac{2}{\delta ^{3/2} (\ell +1)} \langle f_0 \tau_2\rangle_{\delta ,\ell} \,  \mathcal{C}^0(\delta,T) \right) \, . 
\end{align}
where we recall that $I_p=(p_1-p_2)^2+(p_3-p_4)^2$, $J_p=p_1^2+p_2^2+p_3^2+p_4^2$ and
\begin{align}
    P_2&= T \left(\delta  \left(12 \Sigma ^2-44 \Sigma +41\right)+12 \left(\Sigma ^2-4 \Sigma +3\right) T\right) \, , \nonumber \\
    P_1&= \frac{101 \delta }{12}-T J_p-\frac{\delta  I_p}{4}-\frac{1}{2} \delta  (p_2-p_1) (p_3-p_4)+\Sigma ^2 \left(\frac{8 \delta }{3}+6 T\right)+\Sigma  \left(-\frac{28 \delta }{3}-20 T\right)+\frac{37 T}{2} \, ,  \nonumber\\
    P_0&=\frac{1}{2} \Sigma  \left(-2 J_p+2 \left(13+I_p\right) \ell +I_p\right)+J_p+\frac{1}{8} I_p \left(-2 (8 \ell +5)+I_p\right)+\frac{2 \Sigma ^3}{3}-\frac{2 \Sigma }{3}-4 \Sigma ^2 \ell -10 \ell +\frac{5}{8} \, .  \nonumber
\end{align}
For analytic manipulations in the main text, only the averaged OPE data $\langle f_0\rangle$, $\langle \tau_1 \rangle$ and $\langle f_1\rangle$ are needed, together with the constraints \eqref{eq:R4}–\eqref{eq:R1}, which uniquely fix the first curvature correction.

We can solve for the first genuinely new averaged OPE data at this order, namely $\langle f_0\tau_2\rangle$ and $\langle f_2\rangle$, provided in the \texttt{Mathematica} notebook. We also list the leading terms in the main text. A key byproduct of the residue analysis is a simple parity rule
\begin{equation}
\delta-\ell \;\text{ even } \quad\Rightarrow\quad \langle f_0\tau_2\rangle_{\delta,\ell} \;=\; 0 \; .
\label{eq:even-vanish}
\end{equation}
This vanishing follows the behavior of flat-space Veneziano amplitude, where $\langle f_0\rangle_{\delta,\ell} =0 $ when $\delta-\ell$ is even. In the even sector, we will anticipate the OPE data
$\langle f_0\rangle,\langle f_0\tau_1\rangle,\langle f_1\rangle$ and $\langle f_0\tau_2\rangle$ vanish,
but $\langle f_2\rangle$ is different. Importantly, no analogous constraint exists for $\langle f_2\rangle$,
\begin{equation}
\delta-\ell \;\text{ even } \quad\nRightarrow\quad \langle f_2 \rangle_{\delta,\ell}=0 .
\label{eq:f2-even}
\end{equation}
For example, 
\begin{equation}
    \langle f_2 \rangle_{2,0} = (\Sigma -3) (n_t- n_u )+\frac{1}{2} \left(p_1-p_2\right) \left(p_3-p_4\right) \, ,
\end{equation}
is anti-symmetric but non-vanishing under the exchange $t\leftrightarrow u$. The same pattern that even-sector suppression of $\langle f_0\tau_2\rangle$ but not of $\langle f_2\rangle$ is familiar from the $\mathcal{N}=4$ case, where parity selection rules similarly constrain the leading corrections but leave the sub-leading OPE coefficients unconstrained.

\bibliography{aps}
\bibliographystyle{JHEP}

\end{document}